# REPLICATION ATTACK MITIGATIONS FOR STATIC AND MOBILE WSN

V.Manjula[1] and Dr.C.Chellappan [2]

[1]Department of Computer science and Engineering, Anna University, Chennai, India
manjuvv@yahoo.com, drcc@annauniv.edu

## ABSTRACT

*Security is important for many sensor network applications. Wireless Sensor Networks (WSN) are often deployed in hostile environments as static or mobile, where an adversary can physically capture some of the nodes. once a node is captured, adversary collects all the credentials like keys and identity etc. the attacker can re-program it and replicate the node in order to eavesdrop the transmitted messages or compromise the functionality of the network. Identity theft leads to two types attack: clone and sybil. In particularly a harmful attack against sensor networks where one or more node(s) illegitimately claims an identity as replicas is known as the node replication attack. The replication attack can be exceedingly injurious to many important functions of the sensor network such as routing, resource allocation, misbehavior detection, etc.*

*This paper analyzes the threat posed by the replication attack and several novel techniques to detect and defend against the replication attack, and analyzes their effectiveness in both static and mobile WSN.*

## 1. INTRODUCTION

A Wireless Sensor Network (WSN) is a collection of sensors with limited resources that collaborate in order to achieve a common goal. Sensor nodes operate in hostile environments such as battle fields and surveillance zones. Due to their operating nature, WSNs are often unattended, hence prone to several kinds of novel attacks.

The mission-critical nature of sensor network applications implies that any compromise or loss of sensory resource due to a malicious attack launched by the adversary-class can cause significant damage to the entire network. Sensor nodes deployed in a battlefield may have intelligent adversaries operating in their surroundings, intending to subvert damage or hijack messages exchanged in the network. The compromise of a sensor node can lead to greater damage to the network. The resource challenged nature of environments of operation of sensor nodes largely differentiates them from other networks. All security solutions proposed for sensor networks need to operate with minimal energy usage, whilst securing the network. So the basic security requirements of WSN are *availability, confidentiality, integrity and communications* [16].

We classify sensor network attacks into three main categories [7] [8]: Identity Attacks, Routing Attacks & Network Intrusion. Identity attacks intend to steal the identities of legitimate nodes operating in the sensor network. The identity attacks are *Sybil attack and Clone (Replication) attack*. In a Sybil attack, the WSN is subverted by a malicious node which forges a large number of fake identities in order to disrupt the network's protocols. A node replication attack is an attempt by the adversary to add one or more nodes to the network that use the same ID as another node in the network.

Routing attack intend to place the Rogue nodes on a routing path from a source to the base station may attempt to tamper with or discard legitimate data packets. Some of the routing attacks are *Sinkhole Attack, False routing information attack, Selective forwarding attack, and Wormholes*. The adversary creates a large sphere of influence, which will attract all traffic destined for the base station from nodes which may be several hops away from the compromised node which is known as *sinkhole attack*. *False routing attack* means that injecting fake routing control packets into the network. Compromised node may refuse to forward or forward selective packets called as *Selective forwarding attack*. In the *wormhole attack*, two or more malicious colluding nodes create    higher level virtual tunnel in the network, which is employed to transport packets between the tunnel end points.

Network intrusion is an unauthorized access to a system by either an external perpetrator, or by an insider with lesser privileges.

In this paper we are concentrating on an identity attack called replication attack where one or more nodes illegitimately claim an identity of legitimate node and replicated in whole WSN network as shown Figure 1. Reason for choosing this attack is that it can form the basis of a variety attacks such as Sybil attack, routing attacks and link layer attacks etc. also called as denial of service attacks which affects *availability* of network.

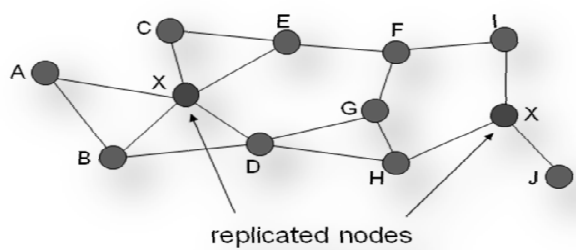

Figure1. Replication Attack

The detection of node replication attacks in a wireless sensor network is therefore a fundamental problem. A few centralized and distributed solutions have recently been proposed. However, these solutions are not satisfactory. First, they are energy and memory demanding: A serious drawback for any protocol that is to be used in resource constrained environment such as a sensor network. Further, they are vulnerable to specific adversary models introduced in this paper.

The rest of this paper is organized as follows; section 2 presents the significance of replication attack nature; section 3 studies analysis of detection and countermeasure of replication attacks and presents discussion and summary. In section 4 presents our proposed model and in section 5 concludes the paper.

## 2. SIGNIFICANCE OF REPLICATION ATTACK AND BACKGROUND

### 2.1 Goals

For a given sensor network, we assume that sensor node not tamper proof and deployed in unattended location. The adversary can capture the node collect all the secret keys, data, and code stored on it. All the credentials are exposed to the attacker. The attacker can easily replicate it in a large number of clones and deploy them on the network. This node replication attack can be the basis for launching a variety of attacks such as DoS attacks and Sybil attacks [7]. If there are many replicated nodes, they can multiply the damage to the network. Therefore, we should quickly detect replicated nodes. The scheme should also revoke the replicated nodes, so that nonfaulty nodes in the network cease to communicate with any nodes injected in this fashion. We evaluate each protocol's security by examining the probability of detecting an attack given that the adversary inserts L replicas of a subverted node. The protocol must provide robust detection even if the adversary captures additional nodes. We also evaluate the efficiency of each protocol. The Communication (for both sending and receiving) among nodes requires at least an order of magnitude power than any other operation. So our first priority to minimize the communication cost for both whole network and individual nodes (hotspots quickly exhausts power), which one of the limitation of WSN. Another limitation is memory. Thus any protocol requiring a large amount of memory will be impractical.

## 2.2 Sensor Network Environments

A sensor network typically consists of hundreds, or even thousands, of small, low-cost nodes distributed over a wide area. The nodes are expected to function in an unsupervised fashion even if new nodes are added, or old nodes disappear (e.g., due to power loss or accidental damage). While some networks include a central location for data collection, many operate in an entirely distributed manner, allowing the operators to retrieve aggregated data from any of the nodes in the network. Furthermore, data collection may only occur at irregular intervals.

For example, many military applications strive to avoid any centralized and fixed points of failure. Instead, data is collected by mobile units (e.g., unmanned aerial units, foot soldiers, etc.) that access the sensor network at unpredictable locations and utilize the first sensor node they encounter as a conduit for the information accumulated by the network. Since these networks often operate in an unsupervised fashion for long periods of time, we would like to detect a node replication attack soon after it occurs. If we wait until the next data collection cycle, the adversary has time to use its presence in the network to corrupt data, decommission legitimate nodes, or otherwise subvert the network's intended purpose.

We also assume that the adversary cannot readily create new IDs for nodes. Newsome et al. describe several techniques to prevent the adversary from deploying nodes with arbitrary IDs. For example, we can tie each node's ID to the unique knowledge it possesses. If the network uses a key predistribution scheme, then a node's ID could correspond to the set of secret keys it shares with its neighbors (e.g., a node's ID is given by the hash of its secret keys). In this system, an adversary gains little advantage by claiming to possess an ID without actually holding the appropriate keys. Assuming the sensor network implements this safeguard, an adversary cannot create a new ID without guessing the appropriate keys (for most systems, this is infeasible), so instead the adversary must capture and clone a legitimate node.

## 3. SOLUTIONS TO REPLICATION ATTACKS AND COUNTERMEASUREMENTS

Solutions to replication attack should follow three key design goals for replica detection schemes. First, replica nodes should be detected with minimal communication, computational, and storage overheads. Second, the detection schemes should be robust and highly resilient against an attacker's attempt to break them. More specifically, the schemes should detect replicas unless the attacker compromises a substantial number of nodes. Finally, there should be no false positives, meaning that only compromised and replica nodes would be detected and revoked. This is important to prevent the attacker from turning a replica detection scheme into a tool for denial of service attacks.

Replication attack detection protocols classified as in the Figure 2 are two categories of static WSN: Centralized and Distributed approaches. These approaches have their own merits and demerits. The main idea of these schemes are to have nodes report location claims that identify their positions and attempt to detect conflicting reports that signal one node in multiple locations. This requires every node to sign and send a location claim, and verify and store the signed location claim of every other node. These protocols, except knowledge about deployment order, are not suitable for mobile WSN since location changes time to time. Figure 3 shows replication attack mitigations in mobile WSN.

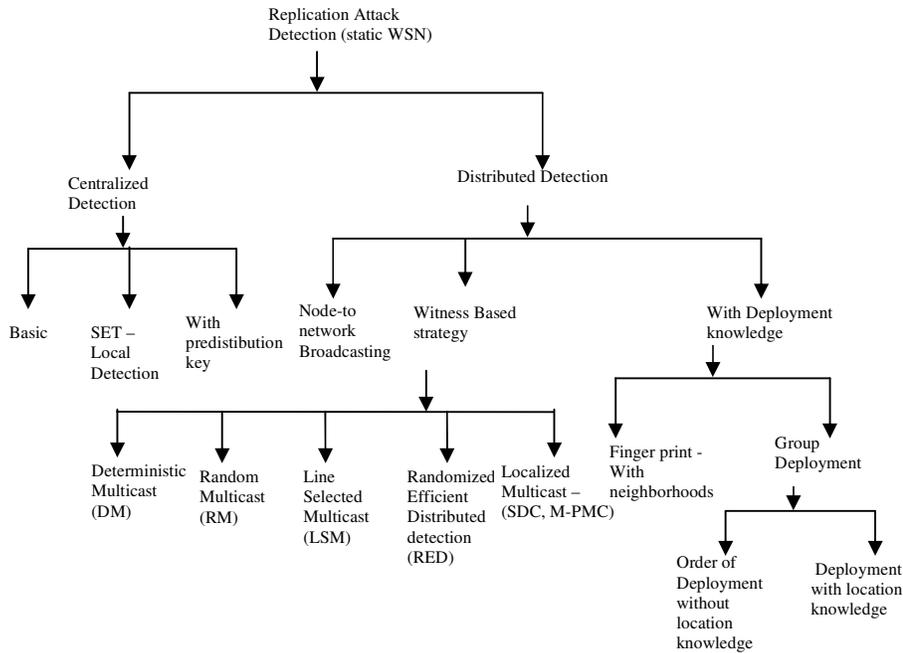

Figure 2. Replication Attack Detection Taxonomy for static WSN

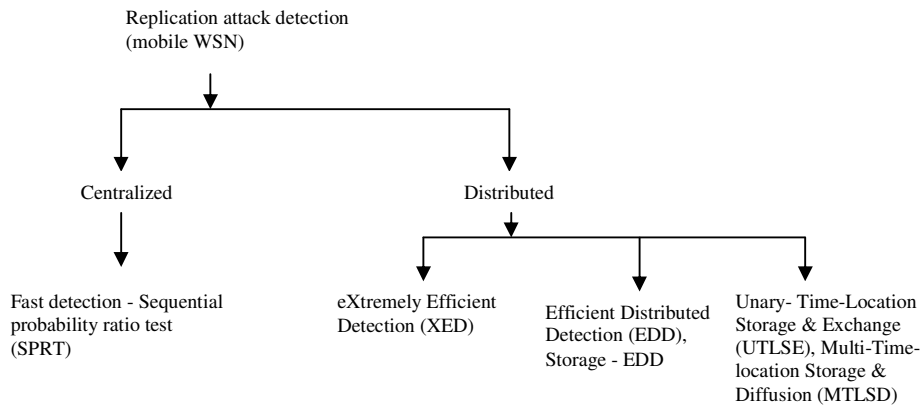

Figure 3. Replication Attack Detection Taxonomy for mobile WSN

## 3.1 Centralized Detection Approaches

In static WSN, The centralized approaches are simple, local detection (SET) and With the Context of random key predistribution and in mobile WSN, fast detection scheme with Sequential probability ratio test (SPRT) have been analysed.

### 3.1.1 Simple Approach

In a simple Centralized approach, the Base Station (BS) acts as centralized entity, each node sends a list of its neighbor nodes and their claimed locations to a base station. If the base station finds that there are two far distant locations for one node ID, then the node clone must have occurred. The BS simply broadcasts through the whole network to expel the cloned nodes. Then, the BS will revoke the replicated nodes. This solution has several drawbacks, for instance: Single point of failure (BS) or any compromise to BS, and high communication cost due to the relevant number of exchanged messages. Furthermore, the nodes closest to the base station will receive the brunt of the routing load and will become attractive targets for the adversary. The protocol also delays revocation, since the base station must wait for all of the reports to come in, analyze them for conflicts and then flood revocations throughout the network. A distributed or local protocol could potentially revoke replicated nodes in a more timely fashion.

### 3.1.2 Local Detection (SET)

Next proposed solutions rely on local detection [4]; using localized voting mechanism, a set of neighbors can agree on the replication of a given node that has been replicated within the neighborhood. However, this kind of method fails to detect replicated nodes that are not within the same neighborhood. SET manages to reduce the communication cost of the preceding approach by computing set operations of exclusive subsets in the network. First, SET launches an exclusive subset maximal independent set (ESMIS) algorithm which forms exclusive unit subsets among one-hop neighbors in an only one disjointed subset which are controlled by a randomly decided leader.

Then those subsets, as in the basic scheme, are transmitted by leaders to the base station such that it can construct all nodes locations and detect clones. Since the subset division procedure eliminates redundancy in the node location reports, SET lowers the communication cost. However, in order to prevent malicious nodes in the ESMIS algorithm, an authenticated subset covering protocol has to be performed, which increases the communication overload and complicates the detection procedure. SET also employs a tree structure to compute non-overlapped set operations and integrates interleaved authentication to prevent unauthorized falsification of subset information during forwarding. Randomization is used to further make the exclusive subset and tree formation unpredictable to an adversary.

### 3.1.3 With the Context of Random Key Predistribution

Brooks et al. [9] propose a clone detection protocol in the context of random key predistribution. The basic idea is that keys that are present on the cloned nodes are detected by looking at how often they are used to authenticate nodes in the network. First each node makes a counting Bloom filter of the keys it uses to communicate with neighboring nodes and appends a nonce. Then Bloom filter and nonce are transferred to base station, which will count the number of times each key is used in the network. Key usage exceeds a threshold can be thought of as suspicious. In fact, it is detecting cloned keys rather than cloned nodes. In the protocol, every node reports its keys to a base station and then the base station uses a statistical approach to find cloned keys. A big problem in this kind of approaches is the high false negative and positive rates. Furthermore, honesty of the malicious nodes while reporting their keys is uncertain.

### 3.1.4 Fast detection with SPRT for Mobile WSN

This section presents the technique to detect replica attacks in mobile sensor networks. In static sensor networks, a sensor node can be considered to be replicated if it is placed at more than one location. However, if nodes are allowed to freely roam throughout the network, the above technique does not work because the mobile node's location will continuously change as it moves. Hence, it is imperative to use some other technique to detect replica nodes in mobile sensor networks. Fortunately, mobility provides us with a clue that can help resolve the mobile replica detection problem. Specifically, a mobile sensor node should never move faster than the system-configured maximum speed [12]. Accordingly, if we observe that the mobile node's speed is over the maximum speed, it is then highly likely that at least two nodes with the same identity are present in the network. We propose a mobile replica detection scheme by leveraging this intuition. It is based on the Sequential Probability Ratio Test (SPRT) which is a statistical decision process. SPRT has been proven to be the best mechanism in terms of the average number of observations that are required to reach a decision among all sequential and non-sequential test processes. SPRT can be thought of as one dimensional random walk with lower and upper limits. Before the random walk starts, null and alternate hypotheses are defined in such a way that the null one is associated with the lower limit and the alternate one is associated with the upper limit. A random walk starts from a point between two limits and moves toward the lower or upper limit in accordance with each observation. If the walk reaches or exceeds the lower or upper limit, it terminates and the null or alternate hypothesis is selected, respectively. We believe that SPRT is well suited for tackling the mobile replica detection problem in the sense that we can construct a random walk with two limits in such a way that each walk is determined by the observed speed of a mobile node; the lower and upper limits are properly configured to be associated with the shortfall and excess of the maximum speed of the mobile node, respectively. We apply SPRT to the mobile replica detection problem as follows. Each time a mobile sensor node moves to a new location, each of its neighbors asks for a signed claim containing its location and time information and decides probabilistically whether to forward the received claim to the base station. The base station computes the speed from every two consecutive claims of a mobile node and performs the SPRT by taking speed as an observed sample. Each time maximum speed is exceeded by the mobile node; it will expedite the random walk to hit or cross the upper limit and thus lead to the base station accepting the alternate hypothesis that the mobile node has been replicated. On the other hand, each time the maximum speed of the mobile node is not reached, it will expedite the random walk to hit or cross the lower limit and thus lead to the base station accepting the null hypothesis that mobile node has not been replicated. Once the base station decides that a mobile node has been replicated, it initiates revocation on the replica nodes.

## 3.2 Distributed Detection Approaches

Distributed detection approaches can be classified broadly in to three categories in Static WSN: *Node-to network Broadcasting, Witness Based strategy, and With Deployment knowledge.* eXtremely Efficient Detection (XED), Efficient Distributed Detection EDD), storage EDD, Unary- Time-Location Storage & Exchange (UTLSE), and Multi-Time- location Storage & Diffusion (MTLSD) are detection approaches in Mobile WSN.

### 3.2.1 Node-to network Broadcasting

This detection approach utilizes a simple broadcast protocol. Basically, each node in the network uses an authenticated broadcast message to flood the network with its location information. Each node stores the location information for its neighbors and if it receives a conflicting claim, revokes the offending node. This protocol achieves 100% detection of all duplicate location claims if the broadcasts reach every node. This assumption becomes false when the adversary jams key areas or otherwise interferes with communication paths through the network. Nodes could employ redundant messages or authenticated acknowledgment techniques to try to thwart such an attack. In terms of efficiency, this protocol requires each node to store location information about its d neighbors. One node's location broadcast requires $O(n)$ messages, assuming the nodes employ a duplicate suppression algorithm in which each node only broadcasts a given message once. Thus, the total communication cost for the protocol is O(n$^2$). Given the simplicity of the scheme and the level of security achieved, this cost may be justifiable for small networks. However, for large networks, the O(n$^2$) factor is too costly, so we investigate schemes with a lower cost.

### 3.2.2 Witness Based strategy

Most of the existing distributed detection protocols [1], [4], [5] adopt the witness finding strategy, in which each node finds a set of sensor nodes somewhere as the witnesses for checking whether there are the same IDs used at different locations, to detect the replicas.

In Deterministic Multicast (DM) [1], to improve on the communication cost of the previous protocol, we describe a detection protocol that only shares a node's location claim with a limited subset of deterministically chosen "witness" nodes. When a node broadcasts its location claim, its neighbors forward that claim to a subset of the nodes called 'witnesses'. The witnesses are chosen as a function of the node's ID. If the Adversary replicates a node, the witnesses will receive two different location claims for the same node ID. The conflicting location claims become evidence to trigger the revocation of the replicated node.

In the Random Multicast (RM) [1], when a node broadcasts its location, each of its neighbors sends (with probability p) a digitally signed copy of the location claim to a set of randomly selected nodes. Assuming there is a replicated node, if every neighbor randomly selects O($\sqrt{n}$) destinations, then exploiting the birthday paradox, there is a non negligible probability at least one node will receive a pair of non coherent location claims. The node that detects the existence of another node in two different locations within the same time-frame will be called witness. The RM protocol implies high communication costs: Each neighbor has to send O($\sqrt{n}$) messages.

In the Line Selected Multicast (LSM)[1]protocol, uses the routing topology of the network to detect replication, each node which forwards claims also saves the claim. That is, the forwarding nodes are also witness nodes of a node which has the node ID in a claim. Therefore, LSM gives a higher detection rate than that of RM. However, both protocols have relatively lower detection rates compared with RED.

In the Randomized Efficient Distributed detection (RED) protocol [5], a trusted entity broadcasts a one-time seed to the whole network. The location of the witness node of a node is determined from the node ID and the seed. Because the seed changes every time, an attacker cannot specify the location of a witness node in advance. The authors of RED said one can also use distributed protocol without a trusted entity such as a local leader election mechanism to create a one-time seed. However, the authors did not mention how to create it; moreover, the local leader election mechanism creates a local leader from a small number of sensor nodes. Even worse, the method does not consider the existence of compromised nodes. Therefore, we cannot use it to create a global leader of a sensor network composed of a large number of nodes with some of them compromised.

In Localized Multicast – (SDC, M-PMC)[4] scheme, each node sends a location claim message to a predetermined cell which is grouped in a geographically separated region. Upon arriving at a cell, this message is broadcasted and stored probabilistically at the witness nodes within the cell. Therefore, the detection rate and the communication overhead are tightly related to the number of nodes and the fraction of witness nodes, which store the location claim message in a cell. However, this scheme is not robust when all nodes within a predetermined cell are compromised.

In the Single Deterministic Cells (SDC) and Parallel Multiple Probabilistic Cells (PMPC) approaches [4], a set of witness nodes located in the vicinity are chosen for each node by using a public known Hash function. Based on the assumption that there is a very efficient way to broadcast a pseudorandom number to all of the sensor nodes periodically, RED [5] also adopts the witness finding strategy to detect the node replication attacks but with less communication cost. The sensor network is considered to be a geographic grid. In the SDC protocol, witness nodes candidates of one node are all nodes of a grid. The grid is statically determined by the node ID, but which nodes in the grid actually become witness nodes are determined randomly. In P-MPC, to increase resiliency to many compromised nodes, the candidate witness nodes for one node are all nodes of several grids.

### 3.2.3 With Deployment knowledge

Bekara and Laurent-Maknavicious proposed a new protocol for securing WSN against nodes replication attacks by limiting the *order of deployment* [9] and no knowledge of nodes deployment locations. Their scheme requires sensors to be deployed progressively in successive generations. Each node belongs to a unique generation. In their scheme, only newly deployed nodes are able to establish pair-wise keys with their neighbors, and all nodes in the network know the number of highest deployed generation. Therefore, the clone nodes will fail to establish pair-wise keys with their neighbors since the clone nodes belong to an old deployed generation.

Xing et al. [10] proposed an approach that achieves real-time detection of clone attacks in WSN. In their approach, each sensor computes a fingerprint by incorporating the neighborhood information through a superimposed s-disjunct code. Each node stores the fingerprint of all neighbors. Whenever a node sends a message, the fingerprint should be included in the message and thus neighbours can verify the fingerprint. The messages sent by clone nodes deployed in other locations will be detected and dropped since the fingerprint does not belong to the same "community".

 Group deployment knowledge scheme [6] is based on the assumption that nodes are deployed in groups, which is realistic for many deployment scenarios. By taking advantage of group deployment knowledge, the proposed schemes perform replica detection in a distributed, efficient, and secure manner. The sensors can be preloaded with relevant knowledge about their own group's membership and all group locations. Then, the sensors in the same group should be deployed at the same time in the location given to that group.  Three schemes have been discussed: Basic, Location claim and multi-group approaches. A basic way to stop replica attack, each node only accept the messages from the member's of their own group(trusted nodes) not from other groups (untrusted  nodes). It stops inter communication between groups. Advantage of this basic scheme is low communication and computational or memory overhead. But the problem is even honest nodes suffers for communication due to deployments points far away from their group. The network becomes poorly connected and not suitable for high resilient applications.  To solve this problem, scheme 2 also forwards messages from untrusted nodes as long as they provide provable evidence that they are not replicas, but based on only predetermined locations for replica detection. Scheme 2 Achieves high replication detection capability with less communication, computational and storage overheads than scheme 1. But there is risk of DoS by flooding fake claims.

To protect against this kind of aggressive adversary, every sensor node sends its neighbor's location claims to multiple groups rather than a single group. This greatly improves our scheme's robustness, while this scheme has higher communication overhead. It can provide a trade-off between the overhead and resilience to attack. This scheme provides very strong resilience to node compromise. Since Attacker needs to compromise multiple groups of nodes to prevent replicas being undetected. Disadvantage of this scheme is more overheads than scheme 2.

### 3.2.4 Mobile WSN Distributed Detection approaches

In *eXtremely Efficient Detection (XED)*[13], the basic operations of this protocol are as follows: Once two sensor nodes encounter each other, they respectively generate a random number, and then exchange the random numbers. If the two nodes meet again, both of them request the other for the random number exchanged at earlier time. If the other cannot reply or replies a number which does not match the number stored in its memory, it announces the detection of a replica. To a smart attacker, this scheme is weak, and he/she can establish secret channels among replicas. By this way, replicas can share the random numbers, and make the protocol fail. Only constant communication cost O(1) is required and the location information of sensor nodes is unnecessary.

The basic idea behind *Efficient and Distributed Detection (EDD) and Storage EDD* [14] schemes is: 1) for network without replicas, the number of times, $\mu_1$, that the node $u$ encounters a specific node $v$, should be limited in a given time interval of length T with high probability 2) for a network with two replicas $v$, the number of times $\mu_2$, that $u$ encounters the replicas with a same ID should be larger than a threshold within the time interval of length T. According to these observations, if each node can discriminate between these two cases, each node has the ability to identify the replicas. The EDD scheme composed of two steps: off-line and on-line. Off-line step is performed by the network planner before sensor deployment, to calculate the parameters time period T and threshold. Online step performed by each node per move. Each checks whether the encountered nodes are replicas by comparing threshold with number of encounter at the end of time interval T. This schemes leads to storage overhead since, each node should maintain list L. To overcome this overhead Storage-efficient EDD scheme proposed. In this SEDD scheme, instead of monitoring all the nodes, each node monitors only subset of nodes called monitor set, in a specific time interval. Storage overhead reduced to cardinality of monitor set.

Two novel *mobility-assisted distributed* [15] solutions to node replication detection in mobile wireless sensor networks are UTLSE and MTLSD. These distributed approach which does not require any routing signaling messages for detecting node replication attacks in mobile wireless sensor networks. The fundamental idea is to make use of the mobility property: Only if two nodes encounter each other, they exchange their time-location claims. That is, if a tracer receives a time-location claim from its tracked neighbor node, it does not immediately transmit this time-location claim to the witness if the witness is not currently within its communication range, but stores that location claim until encountering the witness. In both protocols, after receiving the time-location claims, witnesses carry these claims around the network instead of transmitting them. That means data are forwarded only when appropriate witnesses encounter each other. Unary- Time-Location Storage & Exchange (UTLSE) detects the replicas by each of the two encountered witnesses which stores only one time-location claim. Multi-Time- location Storage & Diffusion (MTLSD), by storing more time-location claims for each tracked node and introducing time-location claims diffusion among witnesses, provides excellent resiliency and sub-optimal detection probability with modest communication overhead. The detection probability of the MTLSD protocol is greater than the probability of protocol UTLSE.

## 4. Summary of Replication attack

Table 1.  Summary of Detection Mechanisms performance overheads

| Schemes | Communication cost | Memory |
|---|---|---|
| SET | $O(n)$ | $O(d)$ |
| Node –to – Network (Broadcast ) | $O(n^2)$ | $O(d)$ |
| Deterministic Multicast | $O(\ g \ln g \sqrt{n}\ / d\ )$ | $O(g)$ |
| Randomized Multicast | $O(n^2)$ | $O(\sqrt{n})$ |
| Line-Selected Multicast (LSM) | $O(n\sqrt{n})$ | $O(\sqrt{n})$ |
| RED | $O(r\ \sqrt{n})$ | $O(r\ )$ |
| SDC | $O(r_f \sqrt{n}) + O(s)$ | $g$ |
| P-MPC | $O(r_f \sqrt{n}) + O(s)$ | g |
| With Deployment Order Knowledge (no location knowledge) | $< O(n\sqrt{n})$ | $< O(\sqrt{n})$ |
| With Neighborhood knowledge - Fingerprint | $O(num_m \sqrt{n})\ \log 2M$ | $O(d) + \min(M , \omega \cdot \log 2\ M)$ |
| With group Deployment Knowledge(basic, location claim, multi-group approach) | $O(m)$ | $O(m)$ |
| | $O(m+d)$ | $O(d+2m)$ |
| | $3O(m+d)$ | $O(d+2m(1+D_{max}\ ))$ |
| XED | $O(1)$ | |
| EDD & SDD | $O(1)\ /O(n)$ | $O(n)\ /O(\xi)$ |
| UTLSE and  MTLSD | $O(n)$ | $O(\sqrt{n})$ |

Where ,

n – No. of nodes in the network

d – Degree of neighboring nodes

g – no. of witness nodes

r- Communication radius

$r_f$ _  No of neighboring nodes forwards location claims

$s$ - The number of sensors in a cell

m –group size

$num_m$ - total number of regular data messages generated during network lifetime

$M$ - the number of rows in the superimposed $s$-disjunct code

$\omega$ - the column weight in the superimposed s-disjunct code.

$D_{max}$ - maximum no. of times that a group servers as the detector group

$\xi$ – Distinct IDs from set of nodes as monitor set

## Conclusion

Table 1 shows the communication cost and storage cost for each technique. In this paper we discussed classification of detection mechanisms for replication attack in static and mobile WSN. Distributed detection approach is more advantages than centralized approaches since single point failure. In witness based strategy of distributed approaches, randomness introduced in choosing witnesses at various levels like whole network and limited to geographical grids to avoid prediction of future witnesses. If chosen witness node itself compromised node or cloned node then detection of replication attack is uncertain. There may be trade-off between memory, communication cost overhead and detection rate. All the approaches dealt with static WSN. With the deployment knowledge (like order, neighborhoods, and group members with locations) all the nodes in the network should know highest deployed generation which impractical and cannot move join other groups since neighbors or fingerprints vary. Some WSN application requires mobile nodes. The entire approaches become complex when considering for mobile nodes which dealt with location claims(only) and Deployment knowledge are not suitable for mobile WSN, since location changes time to time in mobile wireless sensor network. And some other approaches for mobile WSN have been discussed.

## ACKNOWLEDGEMENTS

The authors would like to thank NTRO sponsored Collaborative Directed Basic Research – Smart and Secure Environment Project Lab for providing computing facilities and UGC for financial support by providing fellowship.

**Authors**

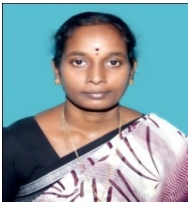

**V.Manjula** received B.E.- Electronics and Communication Engineering (1995)from Thanthai Periyar Govt. Institute of Technology, Vellore,Tamil Nadu, India  under Madras University   and M.E. in Computer Science and Engineering(2000) from Anna University , College Of Engineering  Guindy, Chennai, India. Her current research area is Wireless Sensor Network Security.

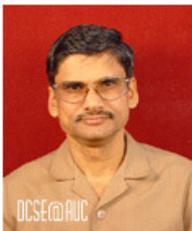

**Dr.C.Chellappan** is a Professor in the Department of Computer Science and Engineering at Anna University, Chennai, India. He received his B.Sc. in Applied Sciences and M.Sc in Applied Science– Applied Mathematics from PSG College Technology, Coimbatore under University of Madras in 1972 and 1977. He received his M.E and Ph.D in Computer Science and Engineering from Anna University in 1982 and 1987. He was the Director of Ramanujam Computing Centre (RCC) for 3 years at Anna University (2002–2005). He has published more than 60 papers in reputed International Journals and Conferences. His research areas are Computer Networks, Distributed/Mobile Computing and Soft Computing, Software Agent, Object Oriented Design and Network Security.